\documentclass[twocolumn,showpacs,preprintnumbers,amsmath,amssymb]{revtex4}

\usepackage{graphicx}
\usepackage{dcolumn}
\usepackage{bm}
\usepackage{color}

\newcommand{\komm}[1]{#1}

\begin{document}


\title{GeV-TeV gamma-rays and neutrinos from the Nova V407 Cygni}

\author{Julian Sitarek}%
 \email{jsitarek@ifae.es}
\affiliation{%
IFAE, Edifici Cn., Campus UAB, E-08193 Bellaterra, Spain \\
on leave from University of \L\'od\'z, Poland
}
\author{W\l odek Bednarek}%
 \email{bednar@astro.phys.uni.lodz.pl}
\affiliation{%
Department of Astrophysics, The University of \L \'od\'z, 90-236 \L \'od\'z, ul. Pomorska 149/153, Poland
}

\date{\today}

\begin{abstract}
The Fermi-LAT telescope has unexpectedly discovered GeV $\gamma$-ray emission from the symbiotic Nova V407 Cygni. 
We investigate the radiation processes due to electrons and hadrons accelerated during the explosion of this Nova.
\komm{We consider a scenario in which} GeV $\gamma$-ray emission observed by Fermi is produced by the electrons with energies of a few tens of GeV in the inverse Compton scattering of stellar radiation.
On the other hand, the hadrons are expected to reach larger energies, due to the lack of radiation losses during acceleration process, producing
TeV $\gamma$-rays and neutrinos. We predict the fluxes of very high energy $\gamma$-rays and neutrinos from Novae of the V407 type for two models of hadron acceleration and discuss their possible detectability by the present and future telescopes (e.g. IceCube, CTA).

\end{abstract}

\maketitle

\section{Introduction}

Nova V407 Cygni belongs to the class of symbiotic Novae which contains a white dwarf and a red giant (RG) star of the Mira-type. 
\komm{The binary period of this system is 43~yrs \citep{mms90}, and the stars are separated by $\sim$10$^{14}$ cm.}
RG has a radius $\sim$500 R$_\odot$, the mass loss rate $3\times 10^{-7}$ M$_\odot$ yr$^{-1}$, the wind velocity $\sim$10 km s$^{-1}$ and the luminosity of $10^4$ L$_\odot$~\komm{\citep{abdo10}}.
The matter of the RG wind is accreted onto the surface of a white dwarf. 
It is commonly accepted that Nova outbursts occur on the surface of a white dwarf as a result of thermonuclear burning of this matter.
It is expected that during the explosion a mass of between a few $10^{-7}$ and  a few $10^{-5}$ M$_\odot$ \citep{or11} has been expelled with a velocity of $2760-3200$ km s$^{-1}$. 
This mass interacts with the wind of RG providing conditions for acceleration of particles.
It is estimated that such symbiotic Novae may appear within the Galaxy with the frequency of $0.5-5$ yr$^{-1}$\citep{lu11}. 
The binary system V407 Cygni is located at a distance of 2.7 kpc \cite{mms90}

Recently, a $\gamma$-ray outburst has been observed from the direction of the binary system V407 Cygni by the Fermi-LAT telescope~\citep{abdo10}. 
The $\gamma$-ray emission was detected during the optical outburst, with the differential spectrum well described in the energy range $0.1-6\,$GeV by a power law with the spectral index $-1.5\pm0.2$ followed by the exponential cutoff at $E_c=2.2\pm0.8\,$GeV.
The emission was detected for $\sim 15$ days (during 2010 March 10-26) with the maximum at 3-4 days after the maximum of the optical emission \citep{nk10}. 
Moreover, the $\gamma$-ray flux at the early stage (10 March - 14 March 2010) and late stage (14 March - 29 March) was different by a factor $\sim$2 while the spectral features remained unchanged within statistical limits~\citep{abdo10}.
V407 Cyg has been also observed by VERITAS at the later phase of the outburst during 2010 March 19-26~\citep{aliu12}. 
Only the upper limit on the flux at the level of $2.3\times 10^{-12}$ erg cm$^{-2}$ s$^{-1}$ has been reported at the energy of 1.6 TeV.   

The detected GeV $\gamma$-ray emission could have been produced in the Inverse Compton (IC) scattering of the RG soft photons by relativistic electrons or as a result of decay of pions produced in hadronic interactions~\citep{abdo10}. 
In fact, modeling of the Novae explosions shows that conditions are convenient for both radiation processes to occur \citep{or11}. Both radiation processes can provide satisfactory fitting of the $\gamma$-ray spectrum measured by the Fermi
~\citep{abdo10}.
If GeV $\gamma$-ray emission is due to the hadronic processes, then neutrinos with $\sim 10\,$GeV energies should be also produced as a result of decay of charged pions and muons. 
It has been argued that such low energy neutrinos (with energies $\sim0.1-100\,$GeV) from symbiotic Novae can be detected by the current and future experiments \citep{raz10}.

In this paper we investigate whether symbiotic Novae can also produce TeV $\gamma$-rays and neutrinos detectable by the present and future instruments. 
We assume that the GeV emission observed by Fermi is due to the IC scattering of the RG radiation by relativistic electrons.
Their maximum energies are determined by radiation energy losses. 
However, hadrons might be accelerated to significantly larger energies due to the lack of saturation of the acceleration process by energy losses. These hadrons might produce TeV $\gamma$-rays and neutrinos in the interaction with the matter of the stellar wind.

\section{General scenario}

\begin{figure}
\vskip 6.4truecm
\includegraphics{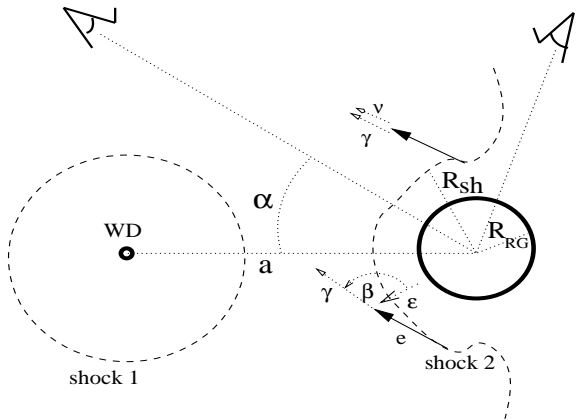}
\caption{Schematic geometry of the symbiotic binary system V407 Cyg. 
The Nova shock wave propagates from the WD star (shock 1) and, as it propagates, starts to partially overtake the RG (shock 2). 
Particles (electrons, hadrons) are accelerated at the shock. 
An electron produces a $\gamma$-ray in IC scattering of a soft photon $\varepsilon$ (angle of interaction $\beta$). 
Hadrons produce $\gamma$-rays and neutrinos in collisions with the matter from RG wind.
}
\label{fig_geometry}
\end{figure}

We apply a standard geometry of symbiotic binary system (see Fig.~\ref{fig_geometry}).
We assume that both electrons and hadrons can be accelerated at the Nova shock wave with comparable luminosities. 
In fact, this does not need to be true since in the case of the shock acceleration scenario the particle Larmor radius has to be larger than the thickness of the shock. 
This condition in general may not be fulfilled by electrons in the blast wave. 
Therefore, in principle electrons and hadrons might be accelerated at different sites and through different acceleration processes. 
For example, the magnetic reconnection, Fermi I and Fermi II processes can be considered here to operate at the shock wave or turbulent medium downstream of the shock.

Taking into account the radiation energy density of RG and the density of matter of its wind (see Introduction) it is clear that the radiation process is expected to be most efficient in the vicinity of the RG.
We assume that the observed GeV $\gamma$-ray emission is produced by electrons in the IC scattering of stellar radiation from the RG.
In fact, a simple estimation of the optical depth for electrons on the IC process (in the Thomson regime) in the radiation field of RG close to its surface gives a value of the order of, $\tau_{\rm IC/T} = n_{\rm RG} R_{\rm RG}\sigma_{\rm T}\sim 3.5$, 
where $n_{\rm RG}$ is the density of stellar photons determined by the surface temperature of this star $T_{\rm RG} = 2.5\times 10^3\,$K$ = 0.25T_4\,$K, $R_{\rm RG} = 3.5\times 10^{13}\,$cm is the radius of RG, and $\sigma_{\rm T}$ is the Thomson cross section. 
This indicates that electrons can lose significant amount of their energy through the production of $\gamma$-rays even on a distance scale comparable to the dimension of the star. In fact, electrons are captured in the vicinity of the star by the wind magnetic field, suffering efficient cooling during the whole lifetime of the flare $\sim$15 days (i.e. corresponding to the 
propagation time of the shock through the separation of the stars in binary system. On the other hand, hadrons can be accelerated in the same region but to much larger energies since their energy loss process is not so efficient as in the case of electrons. These hadrons 
interact with the matter of the wind, losing a part of their energy through pion production (see for estimates \cite{abdo10}). 
$\gamma$-rays and neutrinos are produced as a result of decay of neutral and charged pions with energies which can extend up to TeV energy range.  

\begin{figure}
\includegraphics[width = 0.49\textwidth]{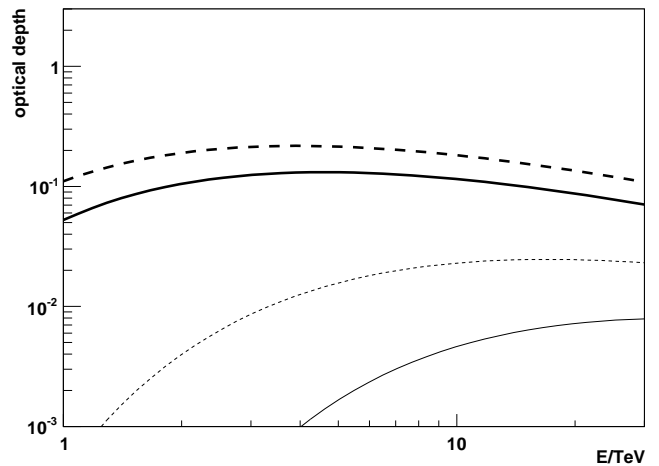}
\caption{Optical depths for $\gamma$-rays escaping from vicinity of the RG star. $\gamma$-rays are produced at the distance of $0.1\,R_{\rm RG}$ (the dashed curves) or $0.5\,R_{\rm RG}$ (the solid curves) from the surface and propagate along the radial direction away from the RG (the thin curves) or perpendicularly to it (the thick lines).}
\label{fig_tau}
\end{figure}

Since the TeV $\gamma$-rays are produced within the radiation field of the RG star, they in principle could be absorbed in the pair production process. However the optical depth for this process is much lower than 1 (see Fig.~\ref{fig_tau}), thus the loss of $\gamma$-rays due to absorption is negligible.

Below we constrain the parameters determining the acceleration mechanism by comparing the $\gamma$-ray spectra produced by electrons in the IC process with the observations of the GeV $\gamma$-ray emission from Nova V407 Cygni by Fermi-LAT. These constraints are farther used to determine the acceleration process of hadrons and predict the expected $\gamma$-ray and neutrino emission from their interaction with the matter of the stellar wind.

\subsection{Constraints on the acceleration of particles}

The gamma-ray spectrum observed from V407 Cygni by Fermi-LAT can be modeled by a power law with an exponential cut-off at energy $E_\gamma\sim 2.2$ GeV \citep{abdo10}. If it is produced in the IC scattering of RG radiation, then in order to produce such $\gamma$-rays, electrons must have energies 
\begin{eqnarray}
E_e\approx m_{\rm e}[{{E_\gamma}\over{2\varepsilon (1+\cos\beta)}}]^{1/2}\approx 
{{21}\over{(1+\cos\beta)^{1/2}}}\ {\rm GeV},
\label{eq1}
\end{eqnarray}
\noindent
where $m_e$ is the electron rest energy, $\varepsilon = 3k_{\rm B}T_{\rm RG} = 2.6T_4$ eV and $\beta$ is the angle between the direction of electron and the thermal photon from the stellar surface. Close to the surface this angle is typically smaller than $0.5\pi$. 

Let us consider the case of saturation of electron acceleration by the IC energy losses in the Thomson (T) regime occurring in the thermal radiation of RG. The cooling time scale of electrons on the IC process in the Thomson regime is estimated on
$\tau_{\rm IC/T} = 3m_{\rm e}^2/4\sigma_{\rm T} c U_{\rm rad}E_{\rm e}\approx 5.4\times 10^4 R_{\rm sh}^2/E_{\rm e}$ s,
where $U_{\rm rad}$ is the energy density of RG radiation and the energy of electrons is expressed in GeV, and $R_{\rm sh}$ is in units of $R_{\rm RG}$. 

The acceleration rate of electrons is parametrized by 
$\dot{P}_{\rm acc} = \xi cE/R_{\rm L}\approx 1.1\xi_{-4}B$ GeV s$^{-1}$,
where $R_{\rm L}$ is the Larmor radius of particles, the energy of electron $E_{\rm e}$ is in GeV, the magnetic field $B$ is in Gauss, $\xi = 10^{-4}\xi_{-4}$ is the acceleration parameter, and $c$ is the speed of light. By comparing the acceleration time scale,
\begin{eqnarray}
\tau_{\rm acc}^{\rm e} = E_{\rm e}/\dot{P}_{\rm acc}\approx 1E_{\rm e}/(\xi_{-4}B)~~~{\rm s}.
\label{eq2}
\end{eqnarray}
with the IC cooling time scale in the T regime, we estimate the maximum energies of accelerated electrons to be (see e.g.~\cite{bp11})
\begin{eqnarray}
E_e^{\rm max}\approx 13(\xi_{-4}B)^{1/2} {{R_{\rm sh}}\over{T_4^2}}\approx 210(\xi_{-4}B)^{1/2} R_{\rm sh}~~~{\rm GeV}.
\label{eq3}
\end{eqnarray}
\noindent 
The value of the acceleration parameter determines the maximum energies of accelerated particles. In the case of the acceleration process occurring in the turbulent medium (Fermi II acceleration process)
it can be estimated on $\xi\sim (v_{\rm turb}/c)^2\sim 10^{-5}$, where $v_{\rm turb}\sim v_{\rm N}/3 = 10^8$ cm s$^{-1}$  are the velocities of the turbulent irregularities produced downstream of the shock, 
$v_{\rm N} = 3\times 10^8v_8$ cm s$^{-1}$ is the velocity of the shock.
We suppose that Fermi II acceleration process is more relevant for electrons which may not be able to fulfill the initial condition for their injection in the shock acceleration scenario (Fermi I acceleration process). However, in the case of hadrons we assume that both mechanism may be relevant.

Note that $E_e^{\rm max}$ should not depend on the distance from the star for the case of a radial structure of the magnetic field around RG. Such structure is expected in the case of a substantial stellar wind in the range of distances characteristic for the separation of the stars in the V407 Cyg binary system.
\komm{Therefore electrons can be injected soon after the Nova explosion.
However the density of radiation from the companion star, which creates a target for IC scattering, increases gradually when the shock approaches the star. 
We expect that the maximum production of $\gamma$-rays by electrons should be determined by the distance to the companion star and the velocity of the shock which might slowly decelerate lowering the acceleration efficiency of electron acceleration.
As a result, the maximum $\gamma$-ray flux produced by electrons in the IC process is expected at some time after optical flash but before the crossing time of the binary separation by the shock, i.e. $\sim$10 days. 
This is in agreement with the measurements of the evolution of the $\gamma$-ray flux measured by Fermi with the maximum $\sim$3-4 days after optical flash.
}

We expect that in the surroundings of the RG star the saturation of the acceleration process of electrons by the synchrotron energy losses is rather unlikely since it requires relatively strong magnetic fields in the stellar wind, $> 40T_4^2$ G, i.e above $B_{\rm crit}\sim 2.5$ G in the case of RG star.
Such large values are not expected in the magnetospheres of horizontal branch of the main sequence solar mass stars.
The simple re-scaling of the surface magnetic field of the Sun ($\sim$3 G) to the dimension of the red giant phase (by a factor $500^{-2}$, assuming conservation of the magnetic field flux, gives the value $\sim$10$^{-5}$ G and $\sim 4\times 10^{-2}$ G for the active regions of its surface (the Suns spots with $B_{\rm spot}\sim 10^4$ G). 
The magnetic dynamo for red giants might be more efficient than the one operating in the main sequence stars but it is rather unlikely that it will be able to produce a magnetic field strength that is larger by a few orders of magnitude.

The comparison of the electron energy constrained by Eq.~\ref{eq1} with that constrained by Eq.~\ref{eq3} allows us to estimate the product of the acceleration efficiency and the magnetic field strength at the shock,
\begin{eqnarray}
\xi_{-4}B\approx 0.015(1+\cos\beta)/R_{\rm sh}^2~~~{\rm G}, 
\label{eq4}
\end{eqnarray}
\noindent
where $R_{\rm sh}$ is in units of $R_{\rm RG}$. 
Note that this condition is satisfied for the magnetic field strengths of the order of $\sim$0.1 G for the required acceleration efficiency of electrons $\xi=10^{-5}$. 
Such a magnetic field strength seems to be reasonable provided that the magnetic dynamo operates more efficiently in the red giants than in the main sequence stars.
Such value of the magnetic field is clearly below the critical value,
$B_{\rm crit}$, above which the synchrotron energy losses of electrons becomes more efficient than their IC losses (note that most of the IC scattering will occur in the Thomson regime). 
Therefore, synchrotron energy losses can be safely neglected in our scenario.

Below we assume that hadrons are also accelerated in the region of electron acceleration characterized by parameters fulfilling the condition
given by Eq.~\ref{eq4}.

\subsection{Maximum energies of accelerated hadrons}

The acceleration time scale of protons, $\tau_{\rm acc}^{\rm p}$, can be derived from an equation analogous to Eq.~\ref{eq2}.
Using the same value of $\xi_{-4}B$ (given by Eq.~\ref{eq4}) as for the acceleration of electrons (i.e. assuming that both populations of particles are accelerated by the same processes in the same region), it can be estimated by,
\begin{eqnarray}
\tau_{\rm acc}^{\rm p} = 1 E_{\rm p}/(\xi_{-4}B)\approx 70R_{\rm sh}^2E_{\rm p}/(1+\cos\beta)~~~{\rm s},
\label{eq5}
\end{eqnarray}
\noindent
where $E_{\rm p}$ is the energy of proton expressed in GeV. 
The typical time scale of the flare (in case of no deceleration of the shock) is
\begin{eqnarray}
\tau_{\rm f} = (a-R_{\rm sh})\times R_{\rm RG} /v_{\rm N}\approx 1.2\times 10^5 
(a-R_{\rm sh})/v_8~~~{\rm s},
\label{eq6}
\end{eqnarray}
where $a\times R_{\rm RG} = 7 R_{\rm RG} = 15.5$ AU is the separation between the stars.
$\tau_{\rm f}$ is of the order of a week (similar to Fermi observations.)
By comparing the acceleration time scale of the protons, given by Eq.~\ref{eq5}, with the  time scale of the flare, given by Eq.~\ref{eq6}, we estimate the maximum energies of accelerated protons by,
\begin{eqnarray}
E_p^{\rm max}\approx 1.7 (a - R_{\rm sh})(1+\cos\beta)/(R_{\rm sh}^2v_8)~~~{\rm TeV}.
\label{eq7}
\end{eqnarray}
For the separation of stars in V407 Cygni given above and the shock arriving to RG surface (i.e. $R_{\rm sh} = 1$), 
the maximum energies of protons are estimated to be $E_{\rm p}^{\rm max}\approx 10$ TeV. 
These relativistic protons interact with the matter of the stellar wind producing charged and neutral pions. 
These products decay to $\gamma$-rays and neutrinos with energies significantly larger than the $\gamma$-rays produced by the GeV electrons which scatter the stellar radiation (estimated above). 
As shown above, the resulting $\gamma$-ray spectrum will not be strongly affected by the absorption in stellar radiation on the pair production process, even if it extends up tens of TeVs.

We conclude that a two component $\gamma$-ray spectrum from the Nova V407 Cyg should be expected in such a model. The lower energy component (in the GeV energy range) is produced in the IC of primary electrons.
The second component (mainly TeV $\gamma$-rays) is produced by hadrons interacting with the wind of RG close to its surface.

\section{Gamma-rays from the Nova V407 Cygni}\label{sec:gamma}

Applying the scenario considered above for the radiation processes, we calculate the IC spectrum produced by electrons which scatter RG radiation when the shock is close to the stellar surface. 
We allow electrons to cool completely in this radiation field.
The energy distribution of electrons is selected such that the produced $\gamma$-ray spectrum is consistent  with the one observed by the Fermi-LAT. 
We consider two models for the electron spectrum: the monoenergetic spectrum (e.g. from magnetic reconnection) and the power law with an exponential cut-off (e.g. Fermi II process).
The second case is defined as below.
Since the cooling time scale of the electrons is clearly shorter than the time scale of the flare (assumed equal to the propagation time of the shock to the RG), we estimate that the injected differential spectrum of electrons should have the form, $dN_{\rm e}/dE_{\rm e}\propto E_{\rm e}^{-1.5}$, in order to produce the $\gamma$-rays in the IC process with the spectral shape $\propto E_\gamma^{-1.75}$.
From the normalization of the calculated $\gamma$-ray spectrum to the observed spectrum, we estimate the total power in the electron spectrum.

We also assume that similar spectral index and total power is characteristic for the spectrum of relativistic protons since they are expected to be accelerated simultaneously by this same acceleration mechanism (note however that they are accelerated to different maximum energies, as their energy loss processes are different). 
The collision time scale of relativistic protons with the matter of the stellar wind (for $p-p\rightarrow \pi$ process) is,
\begin{eqnarray}
\tau_{\rm p-p} = (n_{\rm w}c\sigma_{\rm p-p})^{-1}\approx 10^7n_8^{-1}~~~{\rm s},
\label{eq8}
\end{eqnarray}
\noindent
where the density of matter in the wind is $n = 10^8n_8$ cm$^{-3}$ and the cross section for pion production in collisions of protons in matter is $\sigma_{\rm p-p}\approx 3\times 10^{-26}$ cm$^2$. 
The density of RG wind, close to the stellar surface, is estimated from 
$n = {\dot M}/(4\pi R_{\rm RG}^2m_{\rm p}v_{\rm RG})\approx $\komm{$7.8\times 10^8{\dot M}_{-6.5}/V_6$} cm$^{-3}$,
where $m_{\rm p}$ is the mass of proton, the mass loss rate of the Red Giant is ${\dot M} = $\komm{$3\cdot 10^{-7}{\dot M}_{-6.5}$} M$_\odot$ yr$^{-1}$ and the velocity of the wind from RG is $v_{\rm RG} = $\komm{$10^6V_6$} cm s$^{-1}$.
We conclude that a significant amount of energy (up to a few tens $\%$) can be lost by relativistic hadrons during propagation of the shock through the binary system.

We calculate the $\gamma$-ray spectra from IC scattering of the thermal photons from RG with the help of a dedicated Monte Carlo simulations code.
Since the electrons with energies of a few tens of GeV (close to the cut-off of the spectrum) will no longer interact in the pure Thomson regime, we use the full Klein-Nishina cross-section for the IC scattering in the simulations.
On the other hand the interactions of the relativistic protons with the stellar wind matter are studied with the Geant 4 simulation toolkit \citep{ago03, all06}. 
We apply the QGS model of hadronic interactions which allows us to perform calculations for protons with energies up to 100$\,$TeV.
While multiple hadronic interactions of a single proton are rather rare in the above described scenario (of the order a few \%), they are still automatically taken into account during the simulations.

\begin{figure}
\includegraphics[trim = 0 25 0 0, clip, width = 0.49\textwidth]{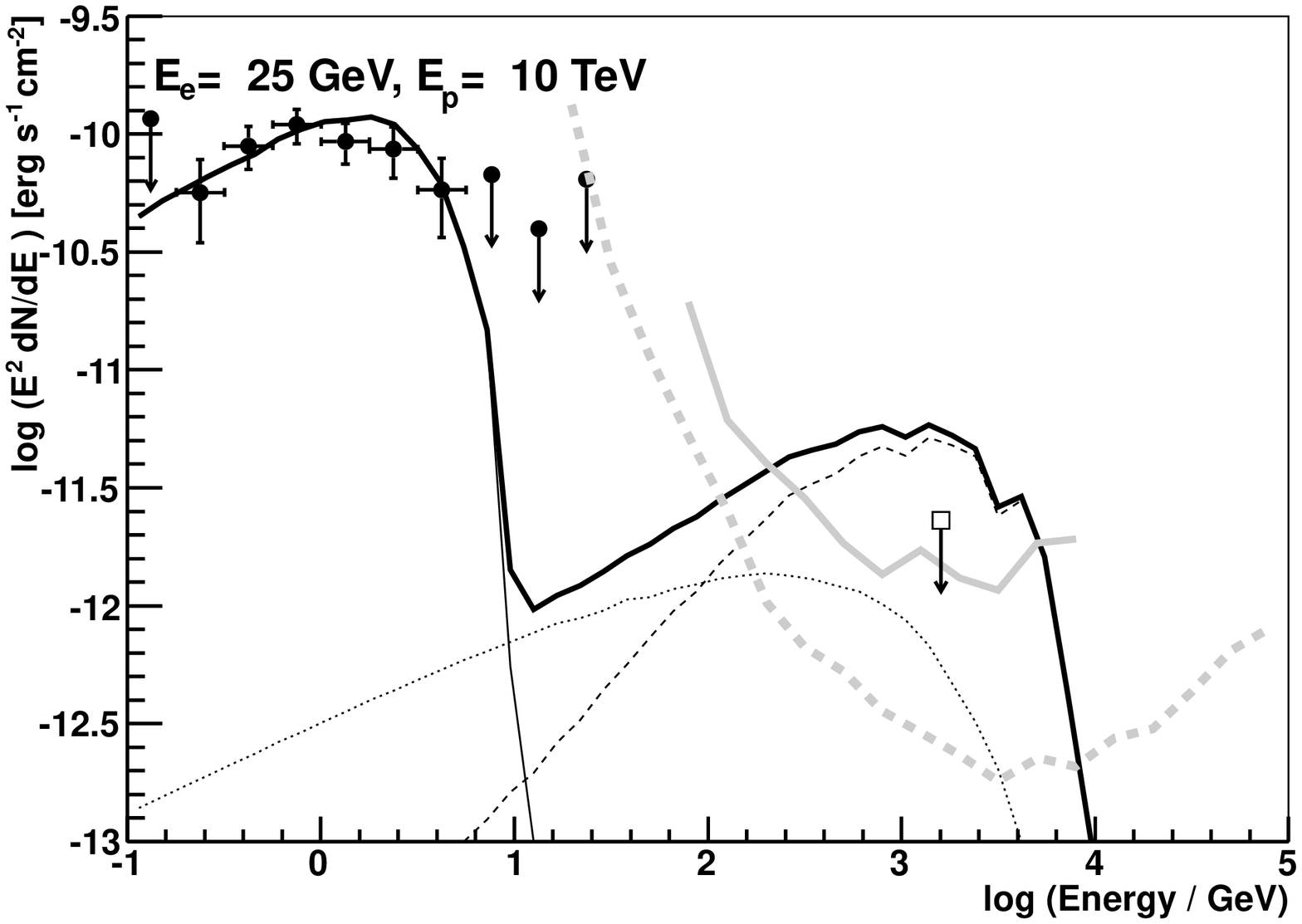}\\
\includegraphics[width = 0.49\textwidth]{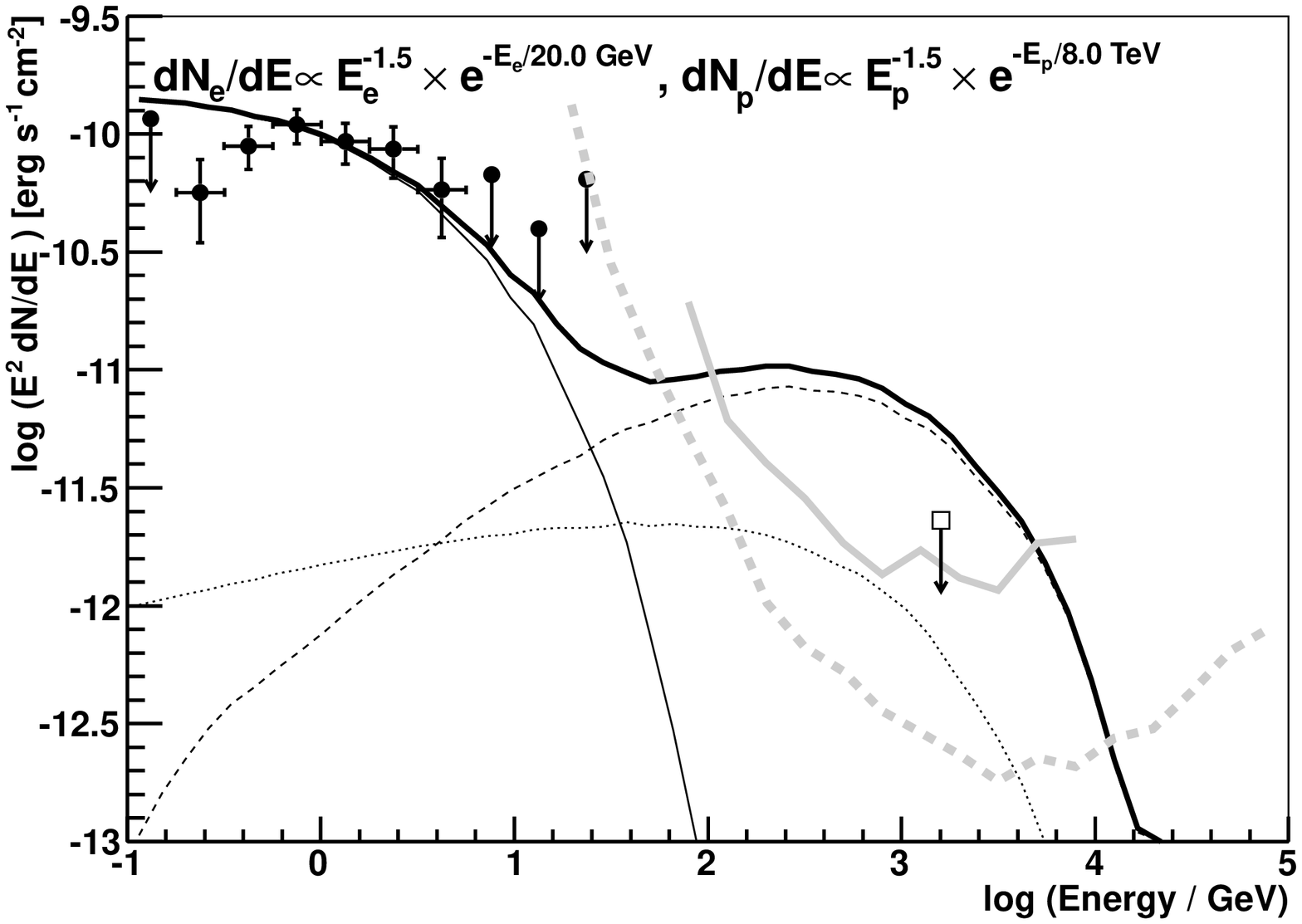}
\caption{
The comparison of the Fermi-LAT $\gamma$-ray average spectral energy distribution (SED) over the period 2010 March 10-29 (black points and upper limits, \citep{abdo10}) with the calculations of the IC $\gamma$-ray spectrum produced by electrons and the TeV $\gamma$-ray spectrum produced in hadronic collisions. 
Monoenergetic electrons with energy $25\,$GeV and protons with energy $10\,$TeV (top panel) or electrons following a spectrum 
$dN/dE_e \propto E_e^{-1.5} \exp(E_e/20\,\mathrm{GeV})$ and protons with spectrum $dN/dE_p \propto E_p^{-1.5} \exp(E_p/8\,\mathrm{TeV})$ (bottom panel) are accelerated in turbulent medium.
Those particles are injected at the distance $0.5\,R_{\rm RG}$ from the surface of RG. The thick black line is the total predicted $\gamma$-ray emission composed of three components:
IC scattering of the injected electrons (the thin solid line), 
$\pi^0$ decay (the thin dashed line) and 
IC scattering of the electrons created in $\pi^\pm$ decay (the dotted line)
Thick gray lines show the differential sensitivity (scaled down to 20 hrs) of the MAGIC telescopes (the solid line, \cite{al12}) and future CTA observatory (dashed line, \cite{actis}).
The empty square marker with an arrow is the flux upper limit from VERITAS observations during the period 2010 March 19-26\citep{aliu12}.
}
\label{fig_gamma}
\end{figure}

As a first example, we consider an acceleration of monoenergetic electrons with energy $\sim 25\,$GeV at the distance $0.5\,R_{\rm RG}$ from the surface of the RG star. 
Simultaneously protons with energy $\sim 10\,$TeV are accelerated at the same shock as predicted by Eq.~7. The electrons will cool down completely in the radiation field of the RG producing GeV $\gamma$-rays. 
We normalize the electron spectrum to reproduce the total $\gamma$-ray flux measured by the Fermi-LAT. 
We assume that the total luminosity of the accelerated protons is equal to that one of electrons. 
We investigate the $\gamma$-ray spectrum obtained from the decay of $\pi^0$ from hadronic interactions.
Moreover, the $e^\pm$ created in the decay of $\pi^\pm$ can also scatter the stellar radiation producing additional component in the $\gamma$-ray spectrum.
Those three components are shown and compared with sensitivities of present and future Cherenkov telescopes in top panel of Fig.~\ref{fig_gamma}.

The luminosity of electrons, $L_e$, required to explain the GeV emission observed by Fermi-LAT is $\sim 4.2 \times 10^{41}\,$erg (assuming they are accelerated over the 14 day period)\komm{, consistent with value $\sim 4 \times 10^{41}\,$erg obtained in modelling of IC emission in \citep{abdo10}}.
With a plausible observation time of 20h, TeV $\gamma$-ray emission from the hadronic component could be detected from a comparable Nova outburst by present Cherenkov telescopes so long as the luminosity of accelerated protons, $L_p$, is above $\sim 0.3$ of the accelerated electron luminosity. 
With the future CTA instrument this component can be detected even if $L_p \gtrsim 0.03 L_e$.
The upper limit on the $\gamma$-ray flux from V407 Cygni derived by the VERITAS telescopes \citep{aliu12} is below the predicted $\gamma$-ray flux for $L_p=L_e$, which is based on the normalization to the average GeV flux observed by Fermi-LAT. 
Simple comparison suggests $L_p \lesssim 0.4 L_e$.
Note however, that the observations by VERITAS were performed 5 days after the peak of the GeV emission. Thus, direct comparison depends on the exact time evolution of $\gamma$-ray emission produced by electrons and hadrons.
A simple investigation of the GeV $\gamma$-ray light curve (Fig.~1 in ~\citep{abdo10}) indicates that the power in relativistic hadrons should be not larger than the power in relativistic electrons assuming similar time evolution of the $\gamma$-ray emission in these two energy ranges.

Secondary pairs, produced in hadronic interactions, have some small effect on the overall spectrum. The IC scattering of these $e^\pm$ pairs softens the total $\gamma$-ray spectrum in the sub-TeV range. It also provides a bridge emission between the low and high energy components. 

As mentioned above, we also consider a scenario in which both, the electrons and the protons, follow a power-law energy distribution with the same spectral index but with different exponential energy cut-offs. 
We assume a spectral index of 1.5 and cut-offs determined by Eq.~\ref{eq2} and Eq.~\ref{eq7}.
The total energy in accelerated electrons $\sim 1.5 \times 10^{42}\,$erg (accelerated over the 14 day period) needed to explain the GeV emission is slightly higher than in the injection of monoenergetic electrons. 
As before we assume that the total luminosities of protons and electrons are comparable. 
The Fermi-LAT spectrum can be fitted for energies above $\sim 0.3\,$GeV, however the predicted flux below this energy is overestimated (see bottom panel of Fig.~\ref{fig_gamma}). 
This may be e.g. due to higher minimal energy of injected electrons, or only partial cooling of the electrons. 
Please note however that measurements in this energy region have high both statistical and systematic errors. 
As the $\gamma$-ray spectrum in sub-TeV energy range is softer in this case than for monoenergetic protons, the effect of the IC scattering of $e^\pm$ produced in the hadronic interactions is less pronounced.
The observations of VERITAS yield a similar limit on the total luminosity of accelerated protons as in the case of monoenergetic injection.
However, since the VERITAS observations were performed with a larger than normal energy threshold (due to high zenith angle of the source during the observations), they did not allow to explore the possible sub-TeV emission, which is predicted to be strong in the power-law injection scenario.

\section{Neutrinos from V407 Cygni}

\begin{figure}[t]
\includegraphics[width = 0.49\textwidth]{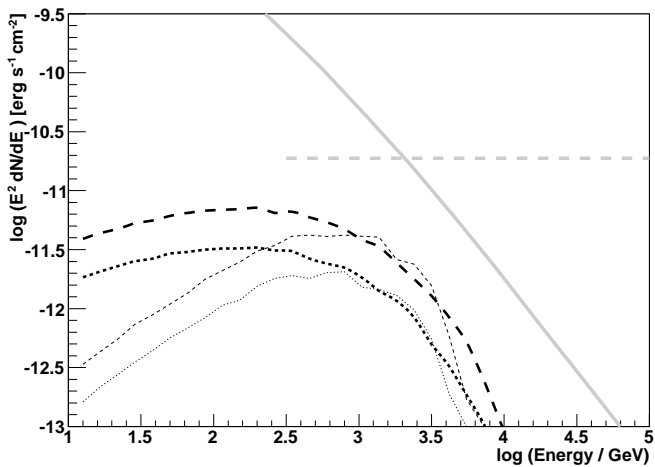}
\caption{
$\nu+\bar{\nu}$ emission from hadronic interaction of $10\,$TeV protons (thin lines) or with spectrum $dN/dE_p \propto E_p^{-1.5} \exp(E_p/8\,\mathrm{TeV})$ (thick lines). 
Black dashed line are the muon neutrinos, and black dotted line are the electron neutrinos. 
For comparison we show the atmospheric muon neutrino flux in a viewcone of $1^\circ$ radius with a thick solid gray line \cite{abb11}, and the sensitivity of the IceCube detector (scaled to 14 day observation time) with a dashed gray line (IceCube design report).
}
\label{fig_nu}
\end{figure}

We consider two models for acceleration of hadrons differing in acceleration efficiency. 
In the first one (described in detail in the previous section), both hadrons and electrons are accelerated either in magnetic reconnection process or the Fermi II acceleration process occurring in the turbulent medium downstream of the shock. 
In the second model, we assume that hadrons are accelerated on the shock wave via the Fermi I process. 
In this case the acceleration efficiency for hadrons (described by the value of $\xi$) is expected to be significantly larger than that for electrons estimated above for the turbulent medium. 
In the Fermi I process $\xi$ can be as large as $\sim v_{\rm N}/c$.

For both proton injection models considered above, we also calculate the neutrino spectra produced in the decay of $\pi^\pm$ (and subsequent $\mu^\pm$). 
In Fig.~\ref{fig_nu}, we show the predicted electron and muon neutrino $\nu+\bar{\nu}$ fluxes for the monoenergetic and the power law injection of TeV protons (see the first example in Section~\ref{sec:gamma}). 
They are compared with the atmospheric neutrino background and the sensitivity of the IceCube telescope. Note that for the considered model the spectra are clearly below these limits for the case $L_{\rm p} = L_{\rm e}$.
As we mentioned above, $L_{\rm p}$ is not likely above $L_{\rm e}$ due to the TeV $\gamma$-ray observations reported by VERITAS~\cite{aliu12}.
Therefore, we conclude that detection of a single outburst by neutrino instruments of the IceCube type is not very likely.

As the second model we consider the case of acceleration of protons with the power law spectrum up to an order of magnitude higher energies than in the previous case and with the luminosity being 10 times larger than that of accelerated electrons.  
Such a scenario might be consistent with the upper limits on the TeV $\gamma$-ray flux derived by VERITAS from observations of V407 Cygni, provided that the time dependence of the acceleration of electrons in the turbulent medium (as considered in the first scenario) and protons on the shock wave in the Fermi I type mechanism also differs significantly.
For example, this might be the case when the magnetic field at the blast wave is dominated by the magnetic field from the White Dwarf. 
Then, the maximum energies of protons are expected to be larger when the magnetic field at the shock is stronger, i.e. at the very early phase of Nova explosion. 
In the latter phase, when the magnetic field at the shock drops, protons are not accelerated to such large energies due to their escape from the shock region. 
On the other hand, electrons start to be efficiently accelerated when the downstream region of the shock becomes more turbulent, i.e. at the phase of the explosion corresponding to the maximum $\gamma$-ray emission observed by Fermi-LAT. 

\begin{figure}
\includegraphics[width = 0.49\textwidth]{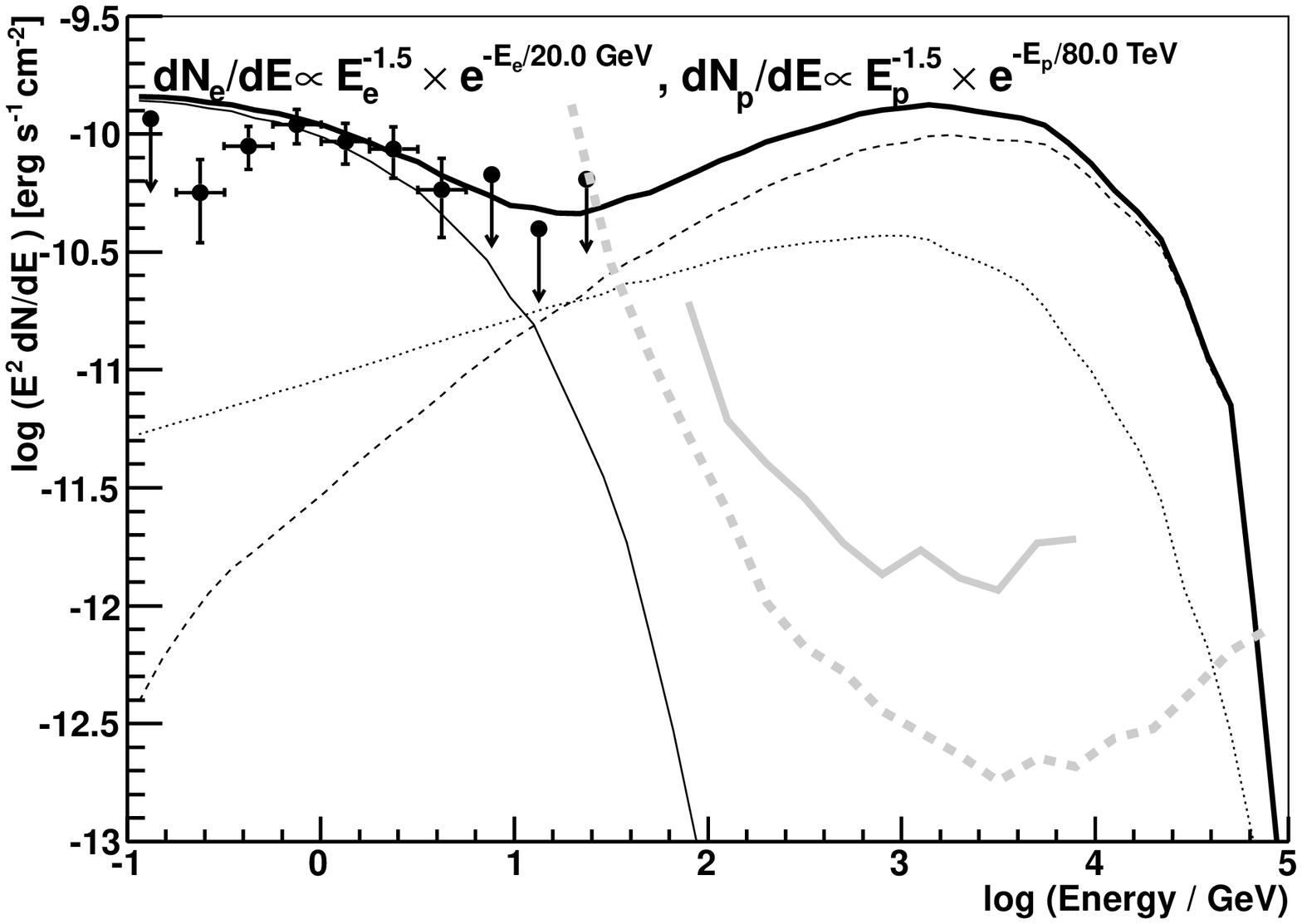}\\
\includegraphics[width = 0.49\textwidth]{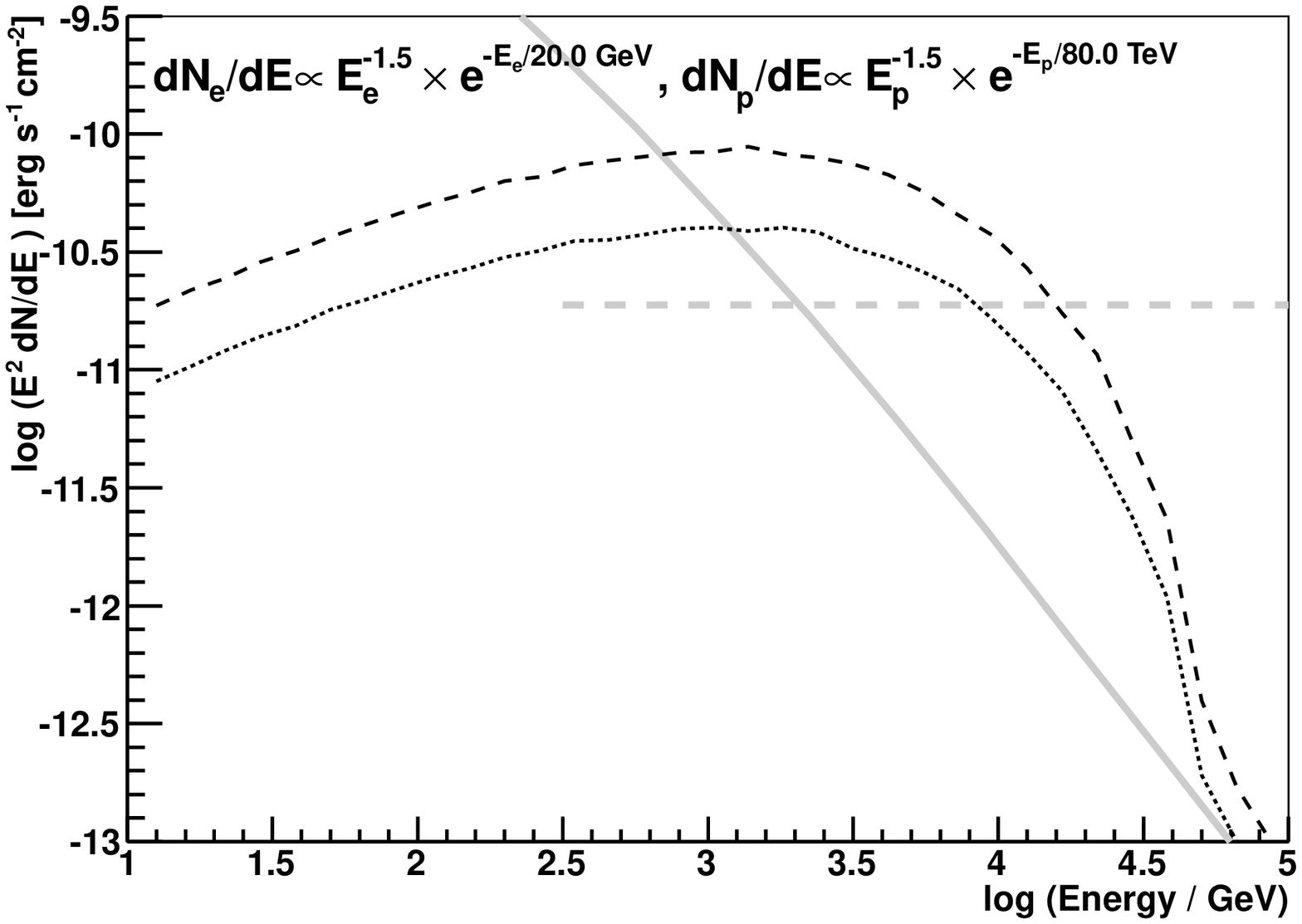}
\caption{SEDs of $\gamma$-rays (the upper panel, like in bottom panel of Fig.~\ref{fig_gamma}) and $\nu+\bar{\nu}$ (the lower panel, like in Fig.~4) but for the case of efficient hadronic acceleration in the first order Fermi acceleration process. 
Injection of electrons $dN/dE \propto E_e^{-1.5} \exp(E_e/20\,\mathrm{GeV})$ 
and protons with spectrum $dN/dE \propto E_p^{-1.5} \exp(E_p/80\,\mathrm{TeV})$
occurs at the distance $0.5\,R_{\rm RG}$ from the surface of RG.
The total power in relativistic protons is 10 times larger than that one in electrons.}
\label{fig_lp10_slope}
\end{figure}

In Fig.~\ref{fig_lp10_slope} we show the expected $\gamma$-ray and neutrino spectra for the case of the power-law acceleration of electrons and protons. 
In this scenario the neutrino spectra overtake the atmospheric background in the multi-TeV range by an order of magnitude. Moreover such a neutrino outburst is above the sensitivity limit of the IceCube telescope. 
However, in this case strong $\gamma$-ray emission from pion decay should be easily detected by the current IACTs even in relatively short time scale. 
The VERITAS upper limit at 1.6 TeV excludes such a possibility during the latter phase of the $\gamma$-ray outburst (i.e. after March 19th 2010).
Therefore, in order to be consistent with those observations hadrons should be efficiently accelerated only during the early phase of the Nova outburst.

\section{Conclusion}

We have considered a two-component scenario for the high energy processes during symbiotic Nova explosions, proposing that GeV $\gamma$-ray emission is produced by electrons and the accompanying TeV $\gamma$-ray and neutrino emission is produced by hadrons. 
\komm{The GeV $\gamma$-ray spectral shape seems to favor monoenergetic electron injection, such as can be obtained e.g. in magnetic reconnection process, rather than power-law electron production typical in e.g. Fermi II acceleration.}
Two models are considered for the acceleration of the hadrons. 
In the first model, hadron acceleration process is tuned to the acceleration process of electrons which is constrained by the GeV $\gamma$-ray observations. 
This model predicts relatively low fluxes of TeV $\gamma$-rays which are marginally consistent with the recent upper limits by VERITAS observations provided that similar power goes on the acceleration of electrons and protons. 
With future CTA project the TeV component in gamma-ray spectrum can be detected even if the luminosity of protons is as low as 3\% of the one of electrons.
The expected neutrino fluxes are below sensitivities of the IceCube telescope. 
In the second model, hadrons are accelerated independently from electrons in a more efficient mechanism (Fermi I for protons in contrast to Fermi II or reconnection for electrons). 
This rather optimistic scenario predicts strong fluxes of TeV $\gamma$-rays which in fact may only appear  during the early stage of the Nova flare (before March 19th) due to the constraints set by the VERITAS upper limits. 
Also the accompanying neutrino burst at TeV energies might be detectable in this case by the IceCube telescope. 

\begin{acknowledgments}
This work is supported by the Polish Na\-ro\-do\-we Cen\-trum Nau\-ki through the grant No.2011/01/B/ST9/00411.
We would like to thank the anonymous referee for the useful comments. 
\end{acknowledgments}


\begin{thebibliography}{}\label{sec:TeXbooks}
\bibitem[Munari et al.(1990)]{mms90} Munari, U., Margoni, R., \& Stagni, R., MNRAS, \textbf{242}, 653 (1990)
\bibitem[Orlando \& Drake(2011)]{or11} Orlando, S., Drake, J.J. MNRAS, \textbf{419}, 2329 (2012)
\bibitem[L\"u et al.(2011)]{lu11} L\"u, G., Zhu, C., Wang, Z., Huo, W., Yang, Y. MNRAS \textbf{413}, L11 (2011).
\bibitem[Abdo et al.(2010)]{abdo10} Abdo, A.A. et al. Science, \textbf{329}, 817 (2010). 
\bibitem[Nishiyama \& Kabashima(2010)]{nk10} Nishiyama, K., Kabashima, F.  IAU Tel:2199 (2010).
\bibitem[Aliu et al.(2012)]{aliu12} Aliu, E., et al. 2012, arXiv:1205.5287 
\bibitem[Razzaque et al.(2010)]{raz10} Razzaque, S., Jean, P., Mena, O. PRD \textbf{82}, 123012 (2010).
\bibitem[Bednarek \& Pabich(2011)]{bp11} Bednarek, W., Pabich, J. A\&A \textbf{530}, 49 (2011).
\bibitem[Agostinelli et al.(2003)]{ago03} Agostinelli, S., et al., NIM A, \textbf{506}, 250 (2003) 
\bibitem[Allison et al.(2006)]{all06} Allison, J., et al., IEEE Transactions on Nuclear Science, \textbf{53}, 270 (2006) 
\bibitem[Aleksi{\'c} et al.(2012)]{al12} Aleksi{\'c}, J., et al., Astropart. Phys., \textbf{35}, 435 (2012) 
\bibitem[Actis et al.(2011)]{actis} Actis, M. et al. Exp. Astron. \textbf{32}, 193 (2011).  
\bibitem[Abbasi et al.(2011)]{abb11} Abbasi, R., et al. , PhRvD, \textbf{83}, 012001 (2011) 
\end{thebibliography}
\end{document}